**Preprint Version**

# Tail Risk Alert Based on Conditional Autoregressive VaR by Regression Quantiles and Machine Learning Algorithms

Nov. 2024

# Tail Risk Alert Based on Conditional Autoregressive VaR by Regression Quantiles and Machine Learning Algorithms

Zong Ke[1st & *]
Faculty of Science
National University of Singapore
Singapore 119077
[*] Corresponding author: a0129009@u.nus.edu

Yuchen Yin[2nd]
Teachers College, Columbia University
525 West 120th Street
New York, NY 10027
yy3243@tc.columbia.edu

*Abstract:* **As the increasing application of AI in finance, this paper will leverage AI algorithms to examine tail risk and develop a model to alter tail risk to promote the stability of US financial markets, and enhance the resilience of the US economy. Specifically, the paper constructs a multivariate multilevel CAViaR model, optimized by gradient descent and genetic algorithm, to study the tail risk spillover between the US stock market, foreign exchange market and credit market. The model is used to provide early warning of related risks in US stocks, US credit bonds, etc. The results show that, by analyzing the direction, magnitude, and pseudo-impulse response of the risk spillover, it is found that the credit market's spillover effect on the stock market and its duration are both greater than the spillover effect of the stock market and the other two markets on credit market, placing credit market in a central position for warning of extreme risks. Its historical information on extreme risks can serve as a predictor of the VaR of other markets.**

***Keywords: genetic algorithm; gradient descent; tail risk; risk warning; financial systems' stability.***

## I. INTRODUCTION

As a random search algorithm, genetic algorithm is derived from biological evolution principles. It simulates natural selection, genetic variation, and crossover, etc. biological processes to search for the optimal solution in the solution space. First, a group of initial solutions is randomly generated, called a population. Then, the fitness function is used to evaluate the goodness of each individual, and individuals with higher fitness have a higher chance of being selected for reproduction. By using the crossover operation, the genetic information of two parental individuals is combined to produce new offspring individuals. At the same time, the mutation operation is introduced, with a certain probability of changing the values of certain genes of an individual to increase the diversity of the population. After many iterations, the population gradually evolves, and excellent individuals continue to emerge, ultimately finding the optimal or approximate optimal solution to the problem.

Gradient descent is a commonly used algorithm in machine learning, which iteratively calculates the gradient of the function and judges the distance between a certain direction and the target, ultimately finding the minimum loss function and related parameters to support the establishment of linear models.

Meanwhile, unlike traditional VaR models, this paper has made improvements and innovations in the following aspects: Firstly, despite the fact that extensive researches have been conducted on the spillover effects of tail risks among different markets, such as commodity, energy and agriculture markets [1-2], the connectedness between stock, bond and foreign exchange markets have not been fully explored. Therefore, based on the multi-variable multi-quantile conditional value-at-risk framework in this paper, the extreme risk levels and spillover effects of major financial markets in the United States have been effectively measured. Unlike mean and volatility spillovers, research on tail risk transmission from a transmission perspective is relatively lacking. Tail risk has destructive power and may lead to systemic risks such as system failure.

Secondly, it is well known, in extreme cases, the conventional patterns between assets will be broken [3-4]. Continuing to use common pattern and data to guide decision-making and risk management in extreme cases at a certain confidence level obviously has many disadvantages and even causes significant losses. One innovation of this article is to study the statistical patterns under tail risk situations, and to provide guidance on how to assist investors in the US financial market in extreme situations by early warning of systemic and tail risks, reducing or even avoiding investment losses in such scenarios, and stabilizing the financial market.

Finally, the transmission direction of extreme financial risks was analyzed using quantitative tools, especially AI algorithms. So far, scholars' discussions on related topics are relatively focused on the degree of correlation, and there is a lack of research on the direction of transmission. Meanwhile, mathematical models in existing researches on contagion mechanism of risks fail to specify the sequential order, path and strength [5-6]. Therefore, this article uses a joint significance test to determine the direction of extreme risk transmission.

## II. LITERATURE REVIEW

Currently, most researchers focus on the conventional correlations of assets. MS-VAR model can reveal the relationship between FX, equity, and fixed income markets evolves according to the valuation of the five advanced countries' currencies [7]. Researchers have found that the empirically observed changes in credit spreads are highly correlated to the dependency among the forex rate and the default risk of the obligor [8-9]. In their research, two different channels are leveraged to capture this dependence: First, default intensities and the diffusion driving FX is probably correlated, and second, when suffered default, one more jump in the exchange rate may appear. The differences between foreign pricing measures and the default intensities under the domestic are studied and closed-form prices for a variety of securities affected by default risk and FX risk are presented (including CDS). Analytical formulas for SPX options and CDX were derived within a structural credit-risk model by jumps and stochastic volatility, as well as leveraging multivariate affine transformation for compound options pricing [10]. The model reveals many aspects of the joint dynamics of SPX options and CDX. However, it cannot reconcile the relative levels of option prices, suggesting that credit and equity markets are not fully integrated [11]. Tail risk in the credit bond market can serve as a warning signal for stock market volatility in China [12]. By analyzing the dynamic conditional correlations between the changes in the general economy's credit risk and stock market returns constructed by the TED spread, researchers found the evidence of significant increases in contemporaneous conditional correlations between changes in the TED spread and stock returns [13]. Also, sterilized support of dollars by the Federal Reserve can weaken the áow of new domestic corporate loans, which has been substantiated by the study on the impact of Forex markets intervention on local credit using a unique central bank dataset of foreign exchange operations [14]. The influence is particularly obvious for banks with thinner capital buyers and for borrowers with larger currency mismatches [15]. Li et al. (2024) proposed to incorporate market sentiment effect into bond yield predictions [16]. Bu et al. (2019) employed deep convolutional network with locality and sparsity constraints for texture classification problems [17], which provided further insights into the application of AI in quantitative finance. Specifically, Wu (2022) proposed Alphanetv4, a deep-learning-based feature extraction and forecast model [18]. Luo (2024) utilized machine learning approach to enhance smart grid efficiency [19]. And Wang et al. (2024) explored into applications of Naïve Bayes classifiers and Bayesian Networks in risk assessment [20].

## III. MODEL

### A. *Model design*

Given that VaR is a discontinuous and nonconvex function for discrete distributions and may cause undesirable results for skewed distributions to cause unreasonable risk management and investment loss, thus, some scholars proposed the conditional autoregressive model of autocorrelation. The conditional auto regressive value at risk (CAViaR model) distracts the attention from the statistical distribution of daily returns directly to the law of the quantile [21]. They specify the movement of the quantile over time with an autoregressive process and take the gradient descent and regression quantile framework suggested by Bassett and Koenker at the same time to determine the estimated parameters.

Here, $q_{t1}(\beta)$ and $q_{t2}(\beta)$ represent two financial markets respectively, and the structured equation captures the risk of two-way linkage between markets.

$$q_{t1}(\beta) = c_1 + a_{11}|Y_{t1-1}| + a_{12}|Y_{t2-1}| + (\beta)q_{t1-1}(\beta) + b_{12}(\beta)q_{t2-1}(\beta) \quad (1)$$

$$q_{t2}(\beta) = c_2 + a_{21}|Y_{t1-1}| + a_{22}|Y_{t2-1}| + b_{21}(\beta)q_{t1-1}(\beta) + b_{22}(\beta)q_{t2-1}(\beta) \quad (2)$$

$$M_1 = \begin{pmatrix} a_{11} & a_{12} \\ a_{21} & a_{21} \end{pmatrix}, M_2 = \begin{pmatrix} b_{11} & b_{12} \\ b_{21} & b_{21} \end{pmatrix} \quad (3)$$

The matrix $M_1$ reflects market shocks, and the coefficient matrix $M_2$ measures tail risk. Among them, b11 (b22) is the autocorrelation coefficient of the tail risk of the financial market, which measures the impact of the previous tail risk on the current tail risk of the current market[22], and b12 (b21) is the tail risk spillover coefficient of the financial market, which measures the impact of the previous tail risk on the current tail risk of another market. The estimation of the MVMQ-CAViaR model can be estimated by the quasi-maximum likelihood estimation method (QMLE[23-24]).

How to estimate parameters? Gradient descent, genetic algorithm and linear regression quantiles claimed by Koenker and Bassett at the end of 1990s are taken and compared at the same time [25]. Gradient descent formula:

$$\beta_{t+1} = \beta_t + r\big(loss(\beta)\big) \quad (4)$$

where r is learning rate, $\beta$ is estimated parameters , $loss(\beta)$ is loss function.

Figure 1 below, shows the framework of genetic algorithm.

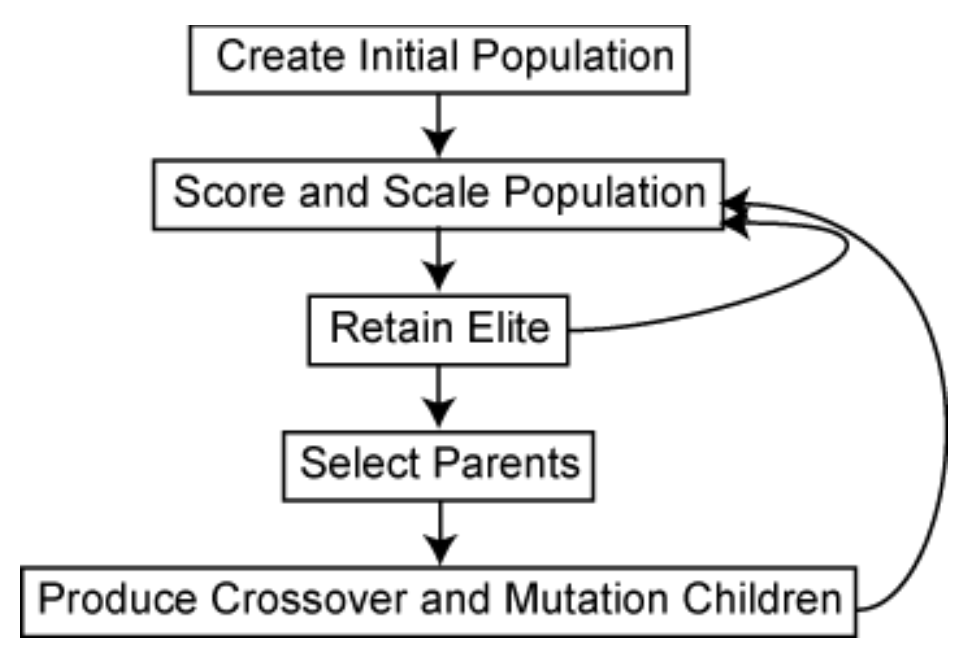

Figure 1 Framework of genetic algorithm

Opt_chromo = 0110011110001110110001010011 (5)

An initial population of 500 single elements or chromosomes is used here to iterate 1000 times. Then, as above, will calculate every single name's fitness score, to evaluate single elements and scale population; after that, will retain elites with crossover, mutation and replacement. Repeat the step again and again to get the optimized results[26-28].

Given a sample of observations $Y_1, \ldots, Y_t$ generated by the following model:

$$Y_t = \beta_i X_t^i + \mu_{kt} \quad Perc(\mu_{kt}|X_t) = 0 \quad (6)$$

here $X_t$ is a multi-vector of regressors and $Perc(\mu_{kt}|X_t)$ is the symbol of k-quantile of $\mu_{kt}$ conditional on $X_t$. Let f (β) = $X_t$β. Then the $k_{th}$ regression quantile is defined as any $\widehat{\beta}$ that solves:

$$\min\left\{ \frac{1}{T} \left[k - i\ (Y_t < f_t(\beta))\right] \left[Y_t - f_t(\beta)\right] \right\} \quad (7)$$

## B. MVMQ-CAViaR model estimation results

### 1) Variable selections and analysis

As shown in table 1 below, ICE BofA US High Yield Index Effective Yield is selected as a proxy indicator for the credit bond market, because the research is tail risk, high-grade credit bonds have lower tail risk and are relatively less sensitive to the market and economy[29-31], and the high-yield market can price most tail risks; the rest, USD Index, Hibor, SPX are selected as proxy variables for the forex market, interbank lending market, and equity market. The daily data of each indicator from May 1, 2014 to June 1, 2024 are selected, and the data that does not match the trading day are eliminated. The daily index yield is expressed by taking the logarithmic transformation of the rate of change of the price of the next period to the price of the previous period, and 2207 research samples are obtained.

Table 1 Variable definitions

| Variables | Symbols | Variable Explanations |
|---|---|---|
| Credits Markets | RCB | ICE BofA US High Yield Index Effective Yield |
| Stock Markets | RS | SPX |
| Interbank Markets | RM | Hibor |
| FX Markets | RE | USD Index |

Table 2 Descriptive statistics of variables

| Variable | Mean | Std | Skewness | kurtosis | JB Statistics | ADF Test | Samples |
|---|---|---|---|---|---|---|---|
| RS | 0.0301 | 1.4026 | -0.4891 | 4.9235 | 401.3269*** | -37.3218 | 2207 |
| RCB | -0.002 | 0.05136 | -0.2794 | 5.7361 | 975.1147*** | -19.0726 | 2207 |
| RM | -0.079 | 8.7521 | 0.2341 | 8.0112 | 2816.3651*** | -41.3261 | 2207 |
| RE | -0.003 | 0.149 | -0.0197 | 4.7261 | 452.0519*** | -53.1176 | 2207 |

Note: *, ** and *** indicate significance at 10%, 5% and 1% levels respectively. The same as in the table below.

According to Table 2, the related fluctuations of the stock market and the inter-bank lending market are significantly ahead of the foreign exchange market and the credit bond market, and their skewness and J-B statistics confirm that they are not standard normal distributions and show certain deviations. Moreover, according to the kurtosis data, their peaks are very obvious, which is in line with the "peaks and thick tails" phenomenon commonly seen in financial markets, with more tail risks[32]. The ADF unit root test shows that at the 1% significance level, each return series is significantly smaller than its critical value, and they are all stationary time series.

### 2) Model estimations

Table 3 shows the risk spillover results at the 1% quantile level. Among them, the stock market, currency market, and foreign exchange market are set as market 1 respectively, and the credit bond market is set as market 2.

Table 3 Estimated results of MCMQ-CAViaR model

| FX markets and Credits markets | | | | | |
|---|---|---|---|---|---|
| Model 1 | c1 | a11 | a12 | b11 | b12 |
| | -0.0057 | -0.3218 | -0.0095 | 0. 1007 | -0.0049 |
| RE/ RCB | (0.0017) * | (0.0419) *** | 0. 0071*** | (0.0007) *** | 0.0101 |
| | c2 | a21 | a22 | b21 | b22 |
| RCB/ RE | -0.0027 | 0. 0081 | -0.0048 | 0.0019 | 1.0108 |
| | (0.0008) *** | -0.039 | 0. 0102*** | -0.0004 | 0.0104*** |
| Stock markets and Credits markets | | | | | |
| Model 2 | c1 | a11 | a12 | b11 | b12 |
| | -0.0901 | -0.1621 | -1.854 | 0. 1021 | -2.647 |
| RS/ RCB | (0.0132)** | (0. 0295)** | (1. 2125) * | (0.0011)*** | (1.1256)** |
| | c2 | a21 | a22 | b21 | b22 |
| RCB/ RS | -0.0394 | -0.0232 | -0.6721 | -0.0103 | 0.0091 |
| | (0. 0114) * | (0.0075)*** | (0.1705)*** | (0.0117)** | (0.0951)*** |
| Interbanks and Credits markets | | | | | |
| Model 3 | c1 | a11 | a12 | b11 | b12 |
| | -0.6182 | -0.756 | 6. 0182 | 0. 638 | 2. 3709 |
| RM/ RCB | (0.1357) ** | (0.1137)*** | (2. 927)** | (0.0072) *** | -1.0947 |
| | c2 | a21 | a22 | b21 | b22 |
| RCB/ RM | -0.012 | 0.1235 | -0.5561 | -0.0054 | 0. 438 |
| | (-0. 0134) | (-0.0106) | (0. 1827)** | (-0. 0218) | 0.0087*** |

#### a) Inter-bank market and credit market

The estimates of the inter-bank lending market and the credit bond market show that coefficients $b_{11}$ and $b_{22}$ are both significant at the 1% confidence level, and the value-at-risk auto-correlation is significant. Coefficients $a_{11}$ and $a_{22}$ are significant at the 1% and 5% confidence levels respectively, indicating that external shocks to the market can significantly affect the value at risk. Coefficient $a_{12}$ is significant but coefficient $a_{21}$ is not significant, indicating that the impact on the credit bond market further deepens the extreme risks of the future inter-bank lending market, while the impact on the inter-bank lending market has no significant impact on the credit bond market.

#### b) Foreign exchange market and credit bond market

Model 3 reports risk spillovers in the foreign exchange market and credit bond market. Same as Model 1 and Model 2, the coefficients $b_{11}$ and $b_{22}$ are both significant at the 1% confidence level, and the two markets have significant auto-correlation; the coefficients $a_{11}$ and $a_{22}$ are significant at the 1% confidence level, which also illustrates the two market has a high level of marketization. Judging from the relationship between the foreign exchange market and the credit bond market, $a_{12}$ is significantly negative at the 10% confidence level, indicating that the impact from the credit bond market will increase the extreme value of risk in the foreign exchange market.

*c) Stock market and credit bond market*

Specifically, the tail risk auto-correlation coefficients of both are significant at the 1% confidence level, and the returns on the stock and bond markets show risk aggregation at the tail. Coefficients $a_{12}$ and $a_{21}$ are the cross-effects of previous shocks on the current market shock, that is, the degree of impact of market information fluctuations in the previous period on the risk value of another market in the current period. They are significantly negative at the 10% and 1% confidence levels respectively. A negative market shock in the previous period will increase the tail risk value of another market, and this effect exists in both the stock market and the bond market. Coefficients $b_{12}$ and $b_{21}$ are the tail risk spillover coefficients of the financial market, both of which are significant at the 5% confidence level, indicating that extreme risks in the stock market can have a significant impact on the credit bond market, and extreme risks in the credit bond market can also be transmitted to stocks.

*3) Optimization using different algorithms*

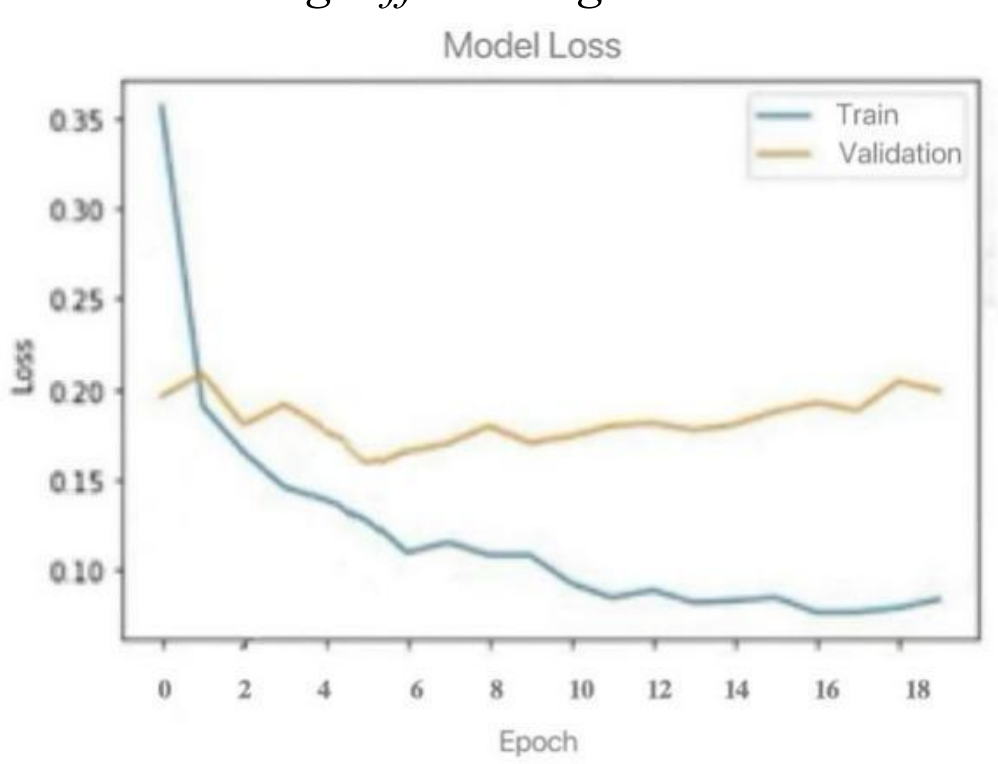

Figure 2 The optimal parameters estimated by gradient descent

As shown in Figure 2, we can see that the model optimized by gradient decent can be converged after enough to some degree. Based on train model, it seems that we should iterate as many as we can, while that validation model shows that iterates to around 5 times are the best option.

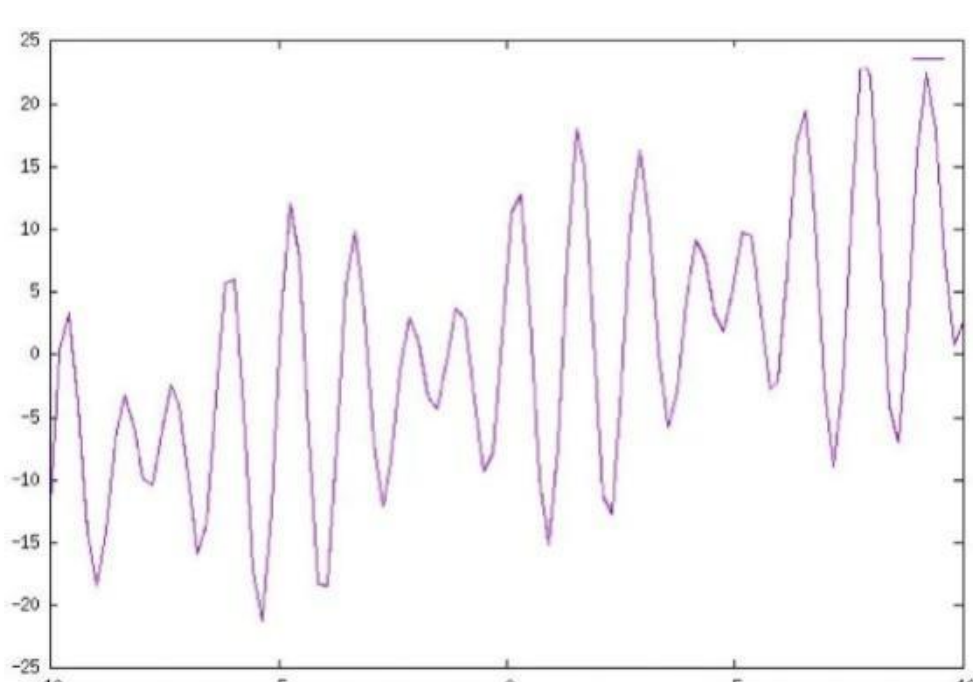

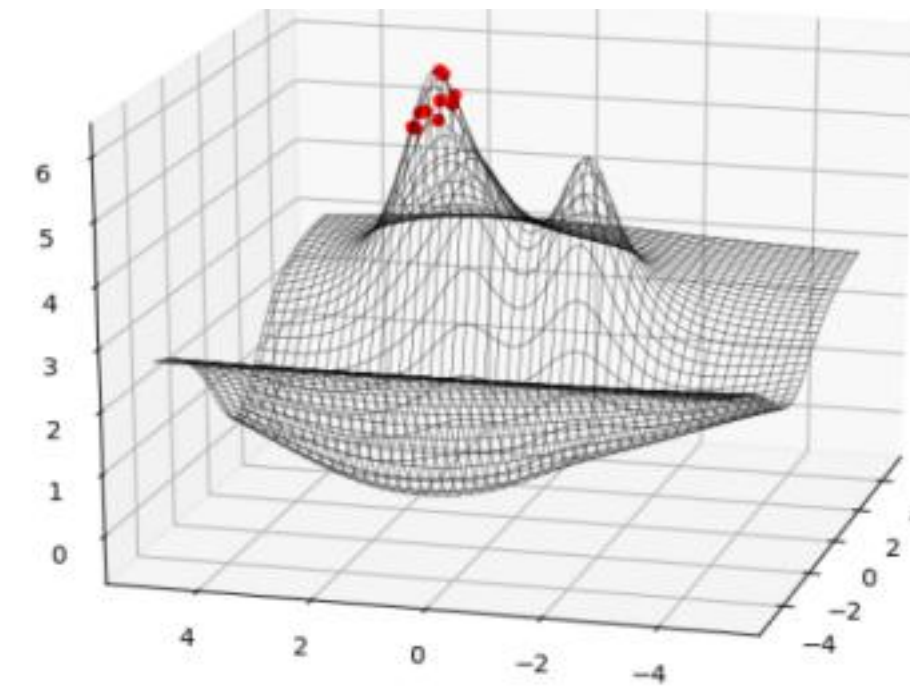

Figure 3 The optimal parameters estimated by genetic algorithms

As shown in Figure 3, optimized by genetic algorithms, the model can be found global optimization solutions. It means that our optimization works pretty well again.

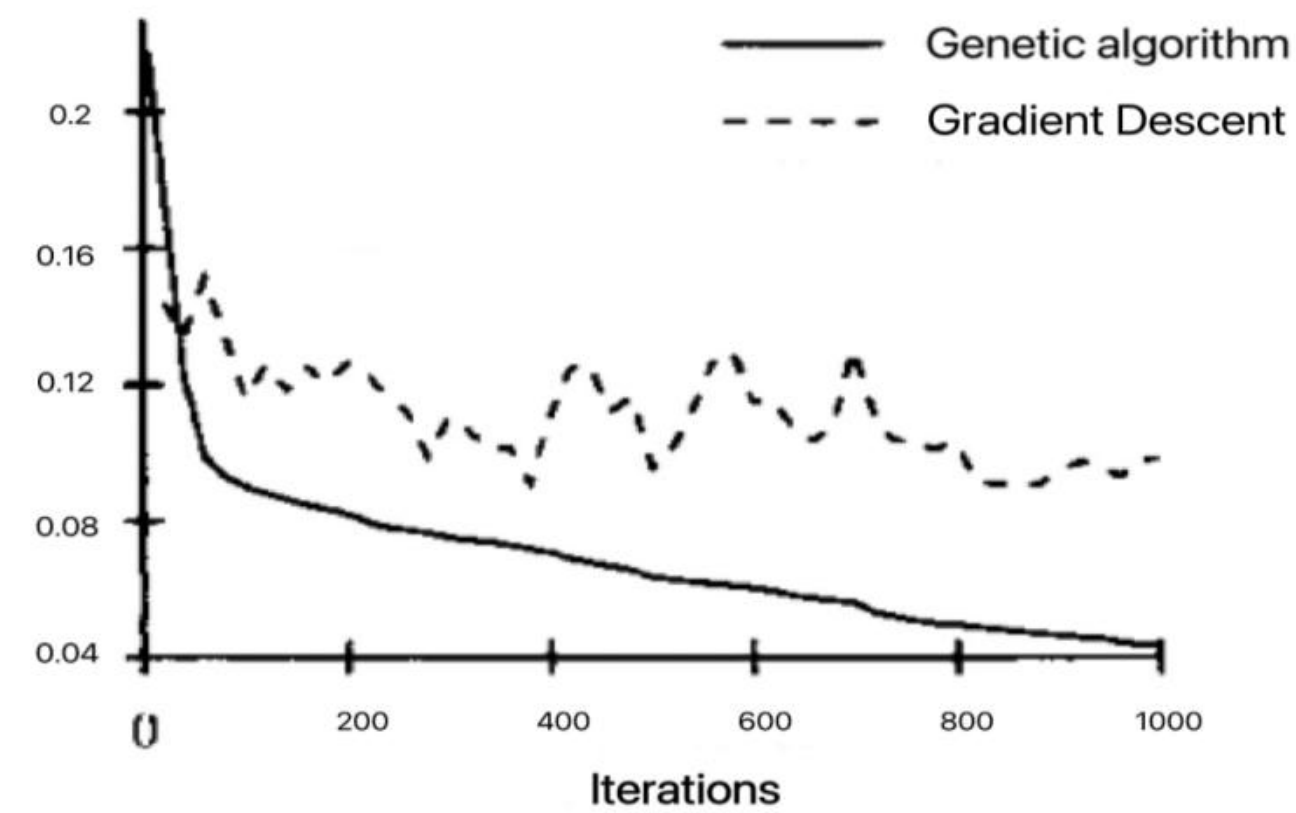

Figure 4 Loss result comparison of different optimization among genetic algorithm and gradient descent

As far as we can see from Figure 4, it compares the optimization results between two different optimization methodologies, i.e., by genetic algorithms and by gradient descent respectively. It can be found that by genetic algorithm, the model will be optimized and converge with the increasing of iterations times, while that by gradient descent, will stop converge after some times of iterations. From optimization perspectives, genetic algorithm looks like a better option for the model in this paper.

*C. Testing of extreme risk spillover effects*

*(1) Analysis of risk spillover*

As mentioned before, extreme risks in financial markets show different rules from risks under statistical significance, and show cross-asset linkage with other markets. Risks in other markets will form spillover effects. In table 4, risk spillovers are fully considered. By constructing joint hypotheses, studying their conduction direction and measure their impact.

Table 4 Joint hypothesis

| Null hypothesis | Risk spillover direction |
|---|---|
| a12= a21 = b12 = b21 = 0 | no extreme risk spillover interaction |
| a21 = b21 = 0 | market 1 no extreme risk spillover interaction market 2 |
| a12 = b12 = 0 | market 2 no extreme risk spillover interaction on market 1 |

Table 5 Joint test results of extreme risk spillover between bond market and other markets

| | Null hypothesis | Chi-square value | P value | Accept/ Refuse |
|---|---|---|---|---|
| Model 1 RE/RCB | a12 = a21 = b12 =b21 = 0 | 8.321 | 0.0619 | refuse |
| | a12 = b12 = 0 | 20.2804 | 0.0004 | refuse |
| | a21 = b21 = 0 | 3.4171 | 0.4906 | refuse |
| Mode 2 RS/RCB | a12 = a21 = b12 =b21 = 0 | 34.3529 | 0.0000 | refuse |
| | a12 = b12 = 0 | 10.0358 | 0.0398 | refuse |
| | a21 = b21 = 0 | 9.1692 | 0.0570 | refuse |
| Mode 3 RM/RCB | a12 = a21 = b12 =b21 = 0 | 31.1218 | 0.0000 | refuse |
| | a12 = b12 = 0 | 27.2192 | 0.0000 | refuse |
| | a21 = b21 = 0 | 3.3862 | 0.4954 | refuse |

As shown in Table 5, each model rejected the null hypothesis, indicating that there is co-movement across markets and, in extreme cases, risk spillovers. In particular, the credit bond market is not only highly correlated with other markets, but also significantly ahead of other markets. Its tail risk fluctuations will spread to other markets. In particular, Model 2 shows that there are two-way spillover effects of extreme risks in the stock market and credit bond market. Models 1 and 3 illustrate that the credit bond market is less affected by the extreme risk spillover effects from the foreign exchange market and the inter-bank lending market, while these two markets are more significantly affected by reverse spillovers from the credit bond market.

(2) Analysis of early warning status in the credit bond market

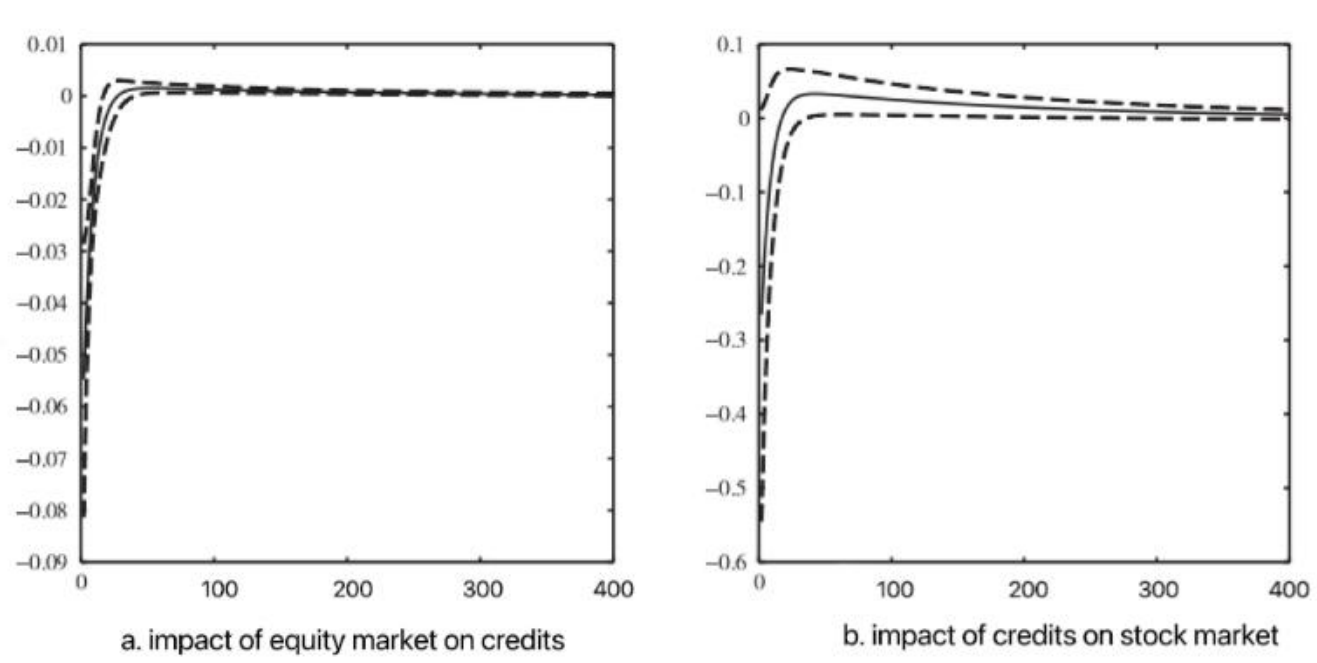

Figure 5 Impacts of extreme risk on credit bond market and stock market

It can be seen from Figure 5 that the credit bond market is only sensitive to extreme risks in the stock market, and is not sensitive to the foreign exchange market and the lending market. However, on the contrary, when the credit bond market encounters tail risks, it will cause obvious impacts and risk spillovers to the other three markets. The effect is obvious, which fully demonstrates that the credit bond market is, to a certain extent, a systemic market as important as the U.S. stock market. Moreover, the pseudo-impulse response analysis found that the spillover effect and duration of the bond market on the stock market are both greater than the spillover effects of the stock market and the other two markets on the credit bond market, thus discovering the importance of risk warning status in the credit bond market.

## IV. Conclusions and Suggestions

This paper uses statistical models and machine learning algorithms to construct a traditional Conditional Autoregressive VaR by Regression Quantiles statistical model, develops a new cross-market risk warning and management analysis tool, improves and measures the accuracy of financial market volatility prediction and risk assessment, focusing on the credit bond market and its tail risk, and its impact on the US stock market, US dollar forex currency market and interbank market, and obtains following research results:

- To some extent, the credit bond market is a systemic market that is equally important as the US stock market. When the credit bond market encounters tail risk, it will have a significant impact on the other three markets and a significant risk spillover effect; extreme risk information from the credit bond market can predict extreme risk changes in other markets and is in a central warning position.
- The impact of the US stock market on the credit bond market is relatively obvious, especially when there is tail risk, the interaction and risk spillover effect of the two markets are obvious.
- However, the risk spillover from the bond market to the stock market is the most obvious, significantly greater than the impact of the stock market on the credit bond market. This shows the importance of the risk warning status of the credit bond market.
- The ups and downs in the forex market and the interbank lending market have little impact on the credit bond market.

In view of the above risk spillover effects, transmission direction, and transmission strength, global investors should take more care of the US credit bond market. At the same time, it is worth trying to establish a credit bond risk warning mechanism based on the method of this article, which is conducive to improving the perception and risk warning of US residents and investment institutions on the credit bond market and the stock market.

To this end, this article puts forward the following suggestions:

- Major financial institutions around the world can moderately increase cross-market risk research, especially the spillover effect, transmission mechanism and risk warning of tail risks.
- Try to use geometric statistical model and genetic algorithms optimization method, to construct a multivariate multi-quantile CAViaR model to study the tail risk spillover between the US credit bond market, the foreign exchange market, and the stock market, and to warn of related risks such as US stock market fluctuations in advance.
- Regulators should promote the interconnection and interoperability of the U.S. stock market, credit bond market, and foreign exchange market, information aggregation and public display, solve the obstacles of retail investors from credit market, learn more about cross-market information, and better manage their position risks.